\documentclass{article}
\usepackage{epsfig}
\usepackage{emulateapj}
\usepackage{apjfonts}

\newcommand{\chandra}{{\it Chandra}}
\newcommand{\asca}{{\it ASCA}}
\newcommand{\rosat}{{\it ROSAT}}

\newcommand{\einstein}{{\it Einstein}}
\newcommand{\lum}{\thinspace\hbox{$\hbox{erg}\thinspace\hbox{s}^{-1}$}}

\newcommand{\snr}{CXOM31\,J004327.7+411829}

\begin{document}

\def\spose#1{\hbox to 0pt{#1\hss}}
\def\laeq{\mathrel{\spose{\lower 3pt\hbox{$\mathchar"218$}}
     \raise 2.0pt\hbox{$\mathchar"13C$}}}
\def\gaeq{\mathrel{\spose{\lower 3pt\hbox{$\mathchar"218$}}
     \raise 2.0pt\hbox{$\mathchar"13E$}}}

\title{The Discovery of a Spatially-Resolved Supernova Remnant in M31
with \chandra}
\author{Albert K.H.~Kong, Michael R.~Garcia, Francis A.~Primini and
Stephen S.~Murray}
\affil{Harvard-Smithsonian Center for Astrophysics, 60
Garden Street, Cambridge, MA 02138; akong@cfa.harvard.edu}

\begin{abstract}
\chandra\ observations of M31 allow the first spatially resolved X-ray
image of a supernova remnant (SNR) in an external spiral galaxy. 
 \snr\ is a slightly elongated ring-shaped object
with a diameter of $\sim 11''$ (42 pc). In addition, the X-ray image
hints that the chemical composition of the SNR is spatial dependent. The X-ray
spectrum of the SNR can be well fitted with a Raymond-Smith model or a
non-equilibrium ionization model. Depending on the spectral model, 
the 0.3--7 keV luminosity
is between $3.2\times10^{36}$ \lum\ and $4.5\times10^{37}$ \lum. The
age of the SNR is estimated to be 3210--22300
years and the number density of ambient gas is $\sim 0.003-0.3$ cm$^{-3}$. This
suggests that the local interstellar medium around the SNR is low.

\end{abstract}


\section{Introduction}

X-ray emission from supernova remnants (SNRs) in our Galaxy has been
extensively studied in the past twenty years. Pre-\chandra\
observations allowed many SNRs to be  resolved, revealing several
different morphologies. The
detailed study of SNR morphology has revealed 
the interaction of ejected materials with
the interstellar medium (ISM) in detail, and also
has provided insights into the evolution of the SNR
progenitor stars.
\chandra\ observations of Galactic SNRs provide exquisite detail,
allowing the chemical stratification of the progenitor to be revealed
(e.g., Canizares et al. 2001) and identifying the central
neutron star and its associated synchrotron nebula (e.g., Slane et al.
2000).
Even with these new insights,  studies of Galactic SNRs can be  
limited by the lack of reliable distance
estimates and  generally high interstellar
absorption. 
These limitations are overcome by studies of 
SNRs in nearby extragalactic systems, the closest of which (the Large
and Small Magellanic Clouds) are near enough to still allow  \chandra\ to reveal 
the structure in detail (see e.g., Hughes 2001). 

Because M31 is the closest ($\sim 800$ kpc) galaxy which shares
similar morphology and size with the Milky Way, it is the best
candidate for a comparative study of SNR. 
Optical surveys for SNR in M31 typically 
identify candidates based on [S\,II] and H$_{\alpha}$ imaging
observations, e.g. 
d'Odorico, Dopita, \& Benvenuti (1980), Blair, Kirshner, \& Chevalier
(1981, 1982), Braun \& Walterbos (1993) and Magnier et al. (1995).
About 200 SNR candidates have been identified by  these surveys,  27
of which have been confirmed via spectroscopic observations (see Blair et
al. 1981, 1982).
In a recent extensive \rosat\ PSPC survey (Supper et al. 2001) of M31, 16
SNRs were identified within a total area of $\sim 10.7$ deg$^2$ by
cross-correlating with the these optical catalogs. Their X-ray
luminosities (0.1--2.4 keV) range from $\sim 10^{36}$ to $\sim
10^{37}$ \lum.

More recently, Kong et al. (2002) detected two SNRs (\snr\
and CXOM31\,J004253.5+412553) in the central $\sim 17'\times17'$
region of M31 with \chandra, both of which were previously
identified by \rosat. It is worth noting that \snr\ was
also identified as a SNR in the \einstein\ data (Blair et
al. 1981). In a recent \chandra\ observation of M31 core region, we
found that one of the SNRs is extended.  In this Letter, we report on
observations of the first SNR in M31 to be resolved at X-ray
wavelengths, \snr .
       
\section{Observations and Data Reduction}

\snr\ was observed several times with \chandra\ Advanced CCD Imaging
Spectrometer (ACIS-I) under a M31
monitoring program during 1999--2001 (see Kong et al. 2002).  \snr\ is
about $8'$ from the aim point of these observations, and therefore the
instrumental point spread function (PSF) is substantially poorer than
if it was on-axis.  In a
recent 5ks ACIS-I observation of a transient in M31 taken on 2001
August 31 (Kong et al. 2001), \snr\ is $< 4'$ from the aim-point,
where the PSF is sufficiently small to allow us to resolve this
object.  Additional observations were obtained with the ACIS-S on 2001
October 5 with total integration time of 37.7ks; although \snr\ is in
the ACIS-S4 front-illuminated (FI) chip and is almost $8'$ from the
aim-point, this long observation provides nearly 300 counts and allows
the best study of the source spectrum. In this Letter, we concentrate
on the analysis of these two latterly mentioned observations; we also
make use of other short ($\sim 5$ks) ACIS observations (see
e.g., Di\,Stefano et al. 2002; Kong et al. 2002) to construct the
long-term lightcurve.

All ACIS data were telemetered in Faint mode, and were collected with
frame transfer time of 3.2 s.  In order to reduce the instrumental
background, we screened the data to allow only photon energies in the
range of 0.3--7 keV, and \asca\ grades of 0, 2, 3, 4, and 6.  In
addition, $\sim 300$\,s of data was excluded from the 2001 October 5
observations due to increased background (likely due to increased
particle flux from the solar wind).  Data were reduced and analyzed
with the \chandra\ X-ray Center CIAO v2.2 package
\footnote{http://asc.harvard.edu/ciao/} and spectral analysis was
performed by making use of XSPEC v11.2
\footnote{http://heasarc.gsfc.nasa.gov/docs/xanadu/xspec/index.html}.
Unless otherwise specified, all quoted errors are at the 1$\sigma$ level. 

\section{Analyses and Results}

\subsection{X-ray Image}

\snr\ was clearly detected in both observations, but we determined the
source position and morphology from the shorter 5ks image because the
source was significantly closer to the aimpoint in this dataset.  We
examined the aspect of this observation with the thread provided by
the CXC
\footnote{http://asc.harvard.edu/mta/ASPECT/fix\_offset/fix\_offset.cgi};
and find an offset of $1.34''$ and $-1.43''$ in
R.A. and Dec., respectively. After correcting for this aspect offset, the
coordinates of the center of \snr\ are found to be 
R.A.=00$^h$\,43$^m$\,27.8$^s$,
DEC.=+41$^{\circ} $18$'$ 29$''$ (J2000) with a positional error of 
about $0.5''$. 
The optical
counterpart of \snr\ is source 2-033 in Magnier et al. (1995),
source 15 in
d'Odorico et al. (1980) and source BA23 in Blair et al. (1981). 
This object was
identified as an irregular, faint and
high [S\,II]/H$_{\alpha}$ ($\sim 1$) SNR. The optical extension is about
$10''-18''$, with a slight elongation along the DEC. direction. 
This source was previously detected (but not resolved)
 with the \rosat\ PSPC and identified
as an SNR (Supper et al. 1997,2001, source RX J0043.4+4118).  It was out of
the field-of-view of the \rosat\ HRI search for SNR in M31 which might have
resolved it (Magnier et al. 1997).  
Figure 1 shows the
``true color'' X-ray image of \snr.  The color scheme is
defined as  0.5--0.8 keV = red, 0.8-1.2 keV = green,  and
1.2--4 keV =  blue. The choice of such energy ranges is to highlight
the O and Ne/Fe emissions (see \S\,3.2) from the SNR.
\snr\ is clearly extended in X-rays as a
distinct ring-shaped object with a
diameter of $\sim 11''$ (42 pc at 780~kpc). The uncertainty of physical dimension of
the SNR is dominated by the distance to M31; for instance, Supper et
al. (1997,2001) used 690 kpc, corresponding to $\sim 37$ pc. Our
quoted distance (780 kpc; Stanek \& Garnavich 1998) is the upper end
of all distance estimates. 

\vspace{2mm}
\vbox{
\centerline{
\psfig{figure=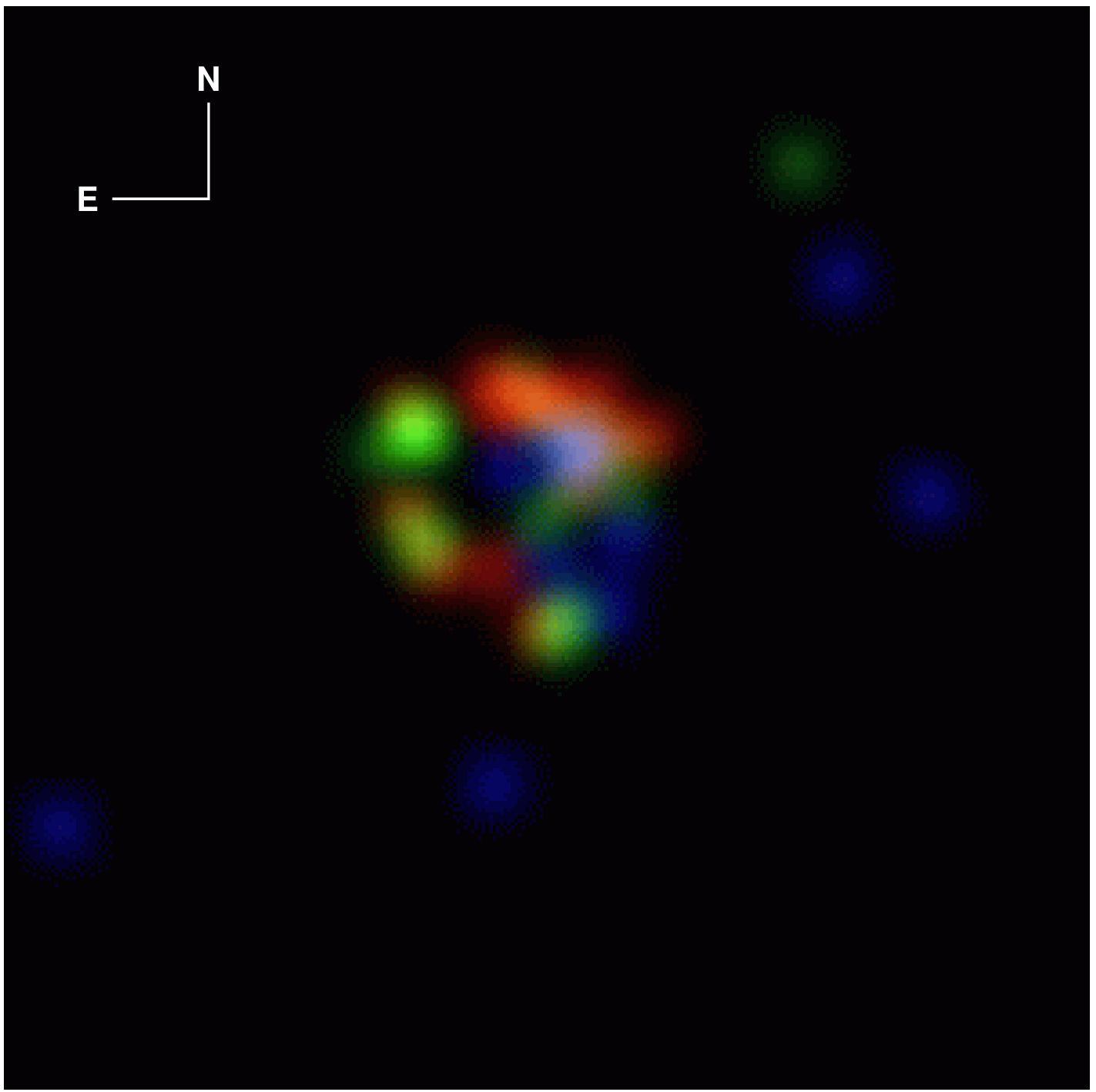,height=3in}
}
\centerline{
\begin{minipage}[h]{3.5in}
{\small Figure 1: {
``True color'' \chandra\ ACIS-I image of
\snr. This image was constructed from the soft (red: 0.5--0.8 keV),
medium (green: 0.8--1.2 keV) and hard (blue: 1.2--4 keV) energy bands. The
pixel size is $0.496''$ and the image has been slightly smoothed with a 
$0.8''\sigma$ Gaussian function. The diameter of the SNR is about
$11''$.}}
\end{minipage}
}}

\subsection{Spectral Analysis}

The long 37.7 ks ACIS observation was used for spectral analysis. We
extracted data from a circle of 12 pixels ($\sim 6''$) radius centered on
\snr\ and background from an annulus with inner and outer radii of 30
and 60 pixels, respectively.  We found 266 net source counts.
In order to allow $\chi^2$ statistics to be used, the spectrum was
grouped into at least 15 counts per spectral bin. Response files were
selected according to the CCD temperature with standard CIAO routines.
We fit the data with several single-component spectral models
including power-law, thermal bremsstrahlung, blackbody,
Raymond-Smith (RS) and
non-equilibrium ionization (NEI) models with interstellar
absorption. The RS model is a simple collisional equilibrium ionization model,
while NEI model is appropriate for modeling SNRs whose age is smaller
than the time required to reach ionization equilibrium. NEI model
consists of electron temperature ($kT_e$) and ionization timescale
($\tau=n_et$), where $n_e$ and $t$ are the mean electron density and the
elapsed time after the plasma was shock heated to a constant
temperature $kT_e$. 
Both models are often
applied to study X-ray
emission from extragalactic SNRs (see e.g., Wang 1999; Schlegel, Blair,
\& Fesen 2000; Hughes, Hayashi, \& Katsuji 1998). 
We were also concerned that the quantum efficiency degradation of ACIS will
affect our spectral fit\footnote{see 
http://asc.harvard.edu/cal/Links/Acis/acis/Cal\_prods/qeDeg/}. The
degradation is shown to be a function of time and is most severe at low
energies. We therefore
applied corrections using the ACISABS absorption model in XSPEC
\footnote{http://www.astro.psu.edu/users/chartas/xcontdir/xcont.html}.
This model allows us to correct the response by inputing the number of
days between {\it Chandra}'s launch and the observation (805 days for
our observation). In general, the degradation mainly affects the best
fit $N_H$, which is over-estimated without the correction.  All
results reported here are corrected for the degradation.

Except for the RS and NEI models, other spectral models give
unacceptable fits to the data ($\chi^2/\nu > 2$). For the RS and NEI
models, we first fixed the abundances at solar value (Andens \&
Grevesse 1989). We also did fits with the abundances of O, N, and Fe
as free parameters and the remaining elements fixed to the solar value.
Finally, we fixed the abundances based on
optical spectroscopy of \snr\ with 
to be $2.3\times10^{-4}$ (O),
$8.4\times10^{-5}$ (N) and $0.72\times10^{-5}$ (S) relative to hydrogen,
respectively (Blair et al. 1982); these correspond to 0.27, 0.75, and
0.44 solar abundance. 
Table 1 lists the
best-fitting parameters for the RS and NEI models. 

\begin{deluxetable}{lcccccccccc}
\tablecaption{Best-fitting Spectral Parameters}
\tablewidth{0pt}
\tablehead{
\colhead{Model} & \colhead{$N_H$} & \colhead{$kT_{RS}/kT_e$} &
\colhead{$\log n_e t$} & \multicolumn{5}{c}{Abundance\tablenotemark{a}} &\colhead{$L_X$}\tablenotemark{b} &
\colhead{$\chi^2_{\nu}$/d.o.f}\\
 &($\times10^{21}$ cm$^{-2})$ &(keV) & & \colhead{N} & \colhead{O} & \colhead{Ne} &
\colhead{S} & \colhead{Fe} & & \\
}
\startdata
RS & $6.78^{+0.49}_{-0.56}$ & $0.14^{+0.03}_{-0.01}$ & & 1 &1&1&1&1&
20.7 & 1.06/13\\
RS & $4.60^{+0.96}_{-2.11}$ & $0.18^{+0.07}_{-0.02}$ & & 1
&$0.49^{+0.39}_{-0.19}$& $0.78^{+0.54}_{-0.26}$ &
 1& $0.14^{+0.12}_{-0.09}$ & 4.5 & 0.89/10\\
RS & $5.88^{+0.50}_{-0.53}$ & $0.14^{+0.03}_{-0.01}$ & & 0.75\tablenotemark{c} & 0.27\tablenotemark{c}&1&0.44\tablenotemark{c}&1&13.4&1.64/13\\
NEI& $0.21^{+0.65}_{-0.21}$ &
$2.1^{+5.6}_{-0.7}$&$10.3^{+0.19}_{-0.16}$& 1 & 1&1&1&1& 0.32 &
0.83/12\\
NEI& $0.25^{+0.62}_{-0.25}$ & $2.1^{+5.6}_{-0.45}$ & $10.3^{+0.19}_{-0.16}$&0.75\tablenotemark{c}&0.27\tablenotemark{c}&1&0.44\tablenotemark{c}&1&0.32&0.83/12\\

\enddata
\tablecomments{All quoted uncertainties are 1$\sigma$. ACISABS model
is applied to correct the degradation of ACIS (see text).}
\tablenotetext{a}{Relative to the solar abundance.}
\tablenotetext{b}{0.3--7 keV luminosity ($\times 10^{37}$\lum),
assuming a distance of 780 kpc; Stanek \& Garnavich 1998.}
\tablenotetext{c}{Fixed at optical value (Blair et al. 1982).}
\end{deluxetable}

All models except the RS model fixed at the optically determined
abundances provide acceptable fits.  The NEI models clearly have 
relatively large errors, especially the electron temperature, $kT_e$.
The best-fit $N_H$ is slightly lower (within 1$\sigma$) than the Galactic value along the
direction ($7\times10^{20}$ cm$^{-2}$; Dickey \& Lockman 1990), and
the lower limit is not well constrained. 
Moreover, the abundance of O, Ne, and Fe is not constrained if they
are left as free parameters in the NEI model.  The best fit
temperatures of the RS models with solar abundances and free O, Ne and
Fe abundances are similar.  The $N_H$ for the models with solar
abundances is an order of magnitude greater than the optically
determined value (see below).  The X-ray spectrum with an absorbed RS model is
shown in Figure 2.  It is evident that
the spectrum is dominated by a broad emission line of O VIII at 0.654
keV and a blend of Fe L shell lines and Ne K shell lines around 0.9
keV. 

\vspace{2mm}
\vbox{
\centerline{
{\rotatebox{-90}{\psfig{file=f2.ps,height=8cm,width=6.5cm}}}
}
\centerline{
\begin{minipage}[h]{3.5in}
{\small Figure 2: {Upper panel: The \chandra\ ACIS-S4 spectrum of \snr\ with
an absorbed RS model ($N_H=4.6\times10^{21}$ cm$^{-2}$, $kT_{RS}=0.18$
keV). Lower panel: Residuals after
subtracting the fit from the data in units of counts s$^{-1}$ keV$^{-1}$.}}
\end{minipage}
}}

\vspace{5mm}
The hydrogen absorption column density for the remnant from optical
observations was estimated to be $N_H=7\times10^{20}$ cm$^{-2}$
($A_V=0.4$; Blair et al. 1981, 1982). The best fit value for $N_H$
from the RS (NEI) model is higher (slightly lower) than the optically determined value. We
also fixed $N_H$ at $7\times10^{20}$ cm$^{-2}$ and the fit is still
acceptable, but gives a slightly higher thermal temperature for both models. 

\subsection{Lightcurve}
The long-term lightcurve is shown in Figure 3, which is constructed
from a series of ACIS pointings (see e.g., Di\,Stefano et al. 2002;
Kong et al. 2001). The background subtracted count rate was corrected
for exposure, background variation and instrumental PSF.  The mean
count rate is 0.007 count s$^{-1}$ 
and it is evident
that the source does not vary significantly as 
 expected for a SNR.

We also tested for variability over shorter time scales during the 
37.7~ks observation. We binned the data into 500 s, 1000
s, 5000 s and 10000 s resolution and used Kolmogorov-Smirnov test to
search for variability, finding nothing significant.

\vspace{1mm}
\vbox{
\centerline{
\psfig{file=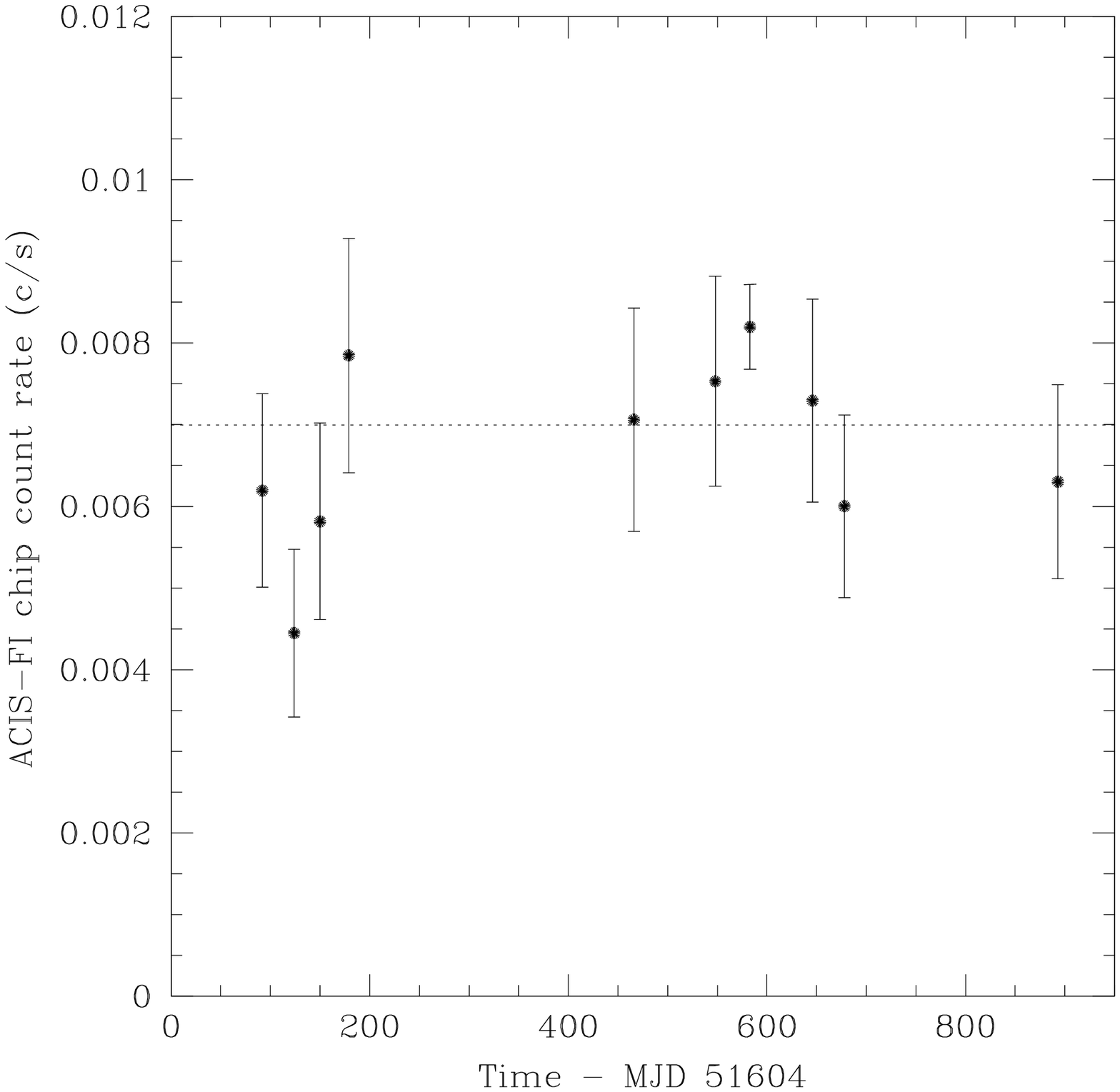,height=8cm,width=8cm}
}
\centerline{
\begin{minipage}[h]{3.5in}
{\small Figure 3: {
Lightcurve of the \chandra\ count rate in the 0.3--7 keV
range of \snr\ since 2000 March. The horizontal line is the average
count rate of the SNR.}}
\end{minipage}
}}

\section{Discussion}

\snr\ is clearly resolved with \chandra\ into a ring-shaped object.
This is the first extragalactic SNR resolved at X-ray wavelengths
(other than those in the Large and Small Magellanic Clouds). The
diameter of \snr\ as measured by \chandra\ is about $11''$,
corresponding to 42 pc; this is comparable to the optically determined
value (d'Odorico et al. 1980; Blair et al. 1981). The X-ray image
shows evidence of a spatial dependence of chemical composition, with O
concentrating in the north-western part, and Ne/Fe distributing along
the eastern and south-eastern arc. It is unclear if the X-ray image
shows the ``blow out'' seen in the
south-eastern section of the optical image (d'Odorico et al. 1980;
Maginer et al. 1995).
There is
no significant variability on both long (months to years) and short
(hours) time scales. The X-ray emission from \snr\ can be acceptably  fit
with a RS model. For abundances fixed at solar values, the estimated
luminosity is near the Eddington luminosity for a $1.4 M_{\odot}$
neutron star, which makes this SNR one of the most luminous ($>
10^{38}$ \lum) objects in M31 (see Kong et al. 2002). Beacaue it is not
realistic to assume solar abundances for all elements, we here consider
the RS model with varing chemical abundances.
Fits to RS model with variable abundances find a slightly lower X-ray
luminosity of $\sim 5\times10^{37}$ \lum\ (0.3--7 keV; $N_H=4.6\times10^{21}$ cm$^{-2}$ and
temperature $kT_{RS}=0.18$ keV), which is
at the high end of luminosity distribution of SNR in M31 (Supper et
al. 1997; Magnier et al. 1997). Assuming
\snr\ is in the adiabatic expansion phase, we can estimate its
physical parameters through the Sedov solution:

\begin{equation}
R=5\,\mbox{pc} \times (E_{51}\,t_3^2/n_0)^{1/5},
\end{equation}

where $R, E_{51}, t_3,$ and $n_0$ are the radius (in pc), initial
explosion energy (in units of $10^{51}$ erg), age (in units of 1000 yr), and
number density of ambient gas (in units of cm$^{-3}$) of the
SNR, respectively. The shock temperature can be written as

\begin{equation}
T_s=(0.18\,\mbox{ keV})(R/t_3)^2,
\end{equation}

Adopting a radius of 21 pc, $T_s=0.18$ keV, and $E_{51}=0.465$ (Blair
et al. 1981); from equations (1) and (2),
We obtain $t=21000$ years and $n_0=0.16$ cm$^{-3}$. Considering the
SNR's size and uncertainty of spectral fit, $t$ ranges between 15700
years and 22300 years while $n_0$ varies between 0.11 cm$^{-3}$ and
0.26 cm$^{-3}$.  We therefore can classify \snr\ as a middle-aged SNR
(cf e.g., Hughes et al. 1998) based on the age derived from the X-ray
spectrum.  Because SNR sometimes contain a pulsar and/or a pulsar
nebula we also did some fits including a power-law component with
photon index fixed at 2 along with our RS model.  These fits limit the
power-law component to less than about 25\% of the flux. From the
derived density, the swept-up mass is estimated to be $(4/3) M_H n_0
\pi R^3 \sim 153 M_{\odot}$, where $M_H$ is mass of a hydrogen
atom. Taking into account of the uncertainty of the SNR's size and
temperature, the swept-up mass ranges from $105 M_{\odot}$ to $172
M_{\odot}$.  Thus for an ejected mass of a few solar masses, the ratio
of swept-up to ejected mass is $\sim 20-30$, indicating that the SNR
is in the Sedov phase.

The X-ray emission from \snr\ can also be fit with an absorbed NEI
model although the uncertainties of $N_H$ and $kT_e$ are large. Using
this model, the
X-ray luminosity (0.3--7 keV) of the SNR is estimated to be $\sim
3\times10^{36}$ \lum, which is typical of SNR in M31. Using Eqns (1)
and (2), we obtain $t=3210-7530$ years and $n_0=0.003-0.02$ cm$^{-3}$. The ratio of
swept-up to ejected mass is $\sim 5$, indicating that the SNR
might be in the early Sedov phase. Comparing
to RS model, these values are smaller. In particular, the number density of ambient gas
is very low. The relatively high $kT_e$ ($=2.1$ keV) is not unusual; for example,
N132D in the Large Magellanic Cloud requires a two-component NEI
model for which the temperatures are $0.8$ and $2.7$ keV
(Favata et al. 1997). The usual explanation for the hot
component is that it comes from the shock-heated swept-up
circumstellar medium, or it is due to the inhomogeneity of the ISM.

The number density of ambient gas
is relatively low ($n_0=0.003-0.26$ cm$^{-3}$). For example, from a
well-studied sample of SNRs in the Large Magellanic Cloud (e.g. Hughes et
al. 1998), $n_0$ ranges between 0.3 to 4 cm$^{-3}$. Previous \rosat\
observations of M31 indicate that the number of X-ray emitting SNRs is
surprisingly small (Supper et al. 1997; Magnier et al. 1997); only 16
out of 200 optically identified SNRs were detected in X-rays.
Magnier et al. (1997) suggested that this lack of X-ray emission
could be due to lower than expected local ISM densities, i.e.,
non-detected SNRs may be in regions with densities 
less than 0.1 cm$^{-3}$, while those SNR detected in 
X-rays may be in higher density regions. 
The low $n_0$ we find for \snr\ provides support for this
suggestion. In addition, in the optical image of Magnier et
al. (1995), the SNR appears to be away from  high density regions of
the galaxy, suggesting that it may be in a relatively low
density region.

Magnier
et al. (1997) note that a multi-component ISM model (McKee \& Ostriker 1977) is
required for M31.  In this case 
the densities we derive herein may be upper limits to the 
dominant low-density component of the ISM in the vicinity of the SNR.  
A future longer observation with the SNR on-axis is required to
investigate multi-component model and detailed morphology.   

\begin{acknowledgements}
We thank Bryan Gaensler and Patrick Slane for discussions, and an
anonymous referee for useful comments. AKHK was
supported by NASA LTSA Grants NAG5-10889 and NAG5-10705.
MRG acknowledges the
support of NASA LTSA Grant NAG5-10889 and NASA Contract NAS8-39073 to
the CXC. The HRC GTO program is supported by NASA Contract
NAS-38248. We acknowledge the hospitality of the Aspen Center for
Physics, which allowed parts of this paper to be written.
\end{acknowledgements}

\end{document}